%% file: main.tex
\documentclass[sigconf]{acmart}

\AtBeginDocument{%
  }

\copyrightyear{2026}
\acmYear{2026}
\setcopyright{cc}
\setcctype{by}
\acmConference[ICSE '26]{2026 IEEE/ACM 48th International Conference on Software Engineering}{April 12--18, 2026}{Rio de Janeiro, Brazil}
\acmBooktitle{2026 IEEE/ACM 48th International Conference on Software Engineering (ICSE '26), April 12--18, 2026, Rio de Janeiro, Brazil}
\acmPrice{}
\acmDOI{10.1145/3744916.3773201}
\acmISBN{979-8-4007-2025-3/2026/04}

\usepackage{multirow}
\usepackage{xcolor}
\usepackage{graphicx}
\usepackage{subcaption}
\usepackage{xspace}
\usepackage{enumitem}
\usepackage{amsmath}

\newcommand{\Ours}{UniCoR\xspace}

\begin{document}

\title{UniCoR: Modality Collaboration for Robust Cross-Language Hybrid Code Retrieval}

\author{Yang Yang}
\affiliation{
  \institution{Central South University}
  \city{Changsha}
  \country{China}
}
\email{yang978@csu.edu.cn}

\author{Li Kuang}
\affiliation{
  \institution{Central South University}
  \city{Changsha}
  \country{China}
}
\email{kuangli@csu.edu.cn}
\authornote{Corresponding author.}

\author{Jiakun Liu}
\affiliation{
  \institution{Harbin Institute of Technology}
  \city{Harbin}
  \country{China}
}
\email{jiakunliu@hit.edu.cn}
\authornotemark[1]

\author{Zhongxin Liu}
\affiliation{
  \institution{Zhejiang University}
  \city{Hangzhou}
  \country{China}
}
\email{liu_zx@zju.edu.cn}

\author{Yingjie Xia}
\affiliation{
  \institution{Hangzhou Dianzi University}
  \city{Hangzhou}
  \country{China}
}
\email{xiayingjie@zju.edu.cn}

\author{David Lo}
\affiliation{
  \institution{Singapore Management University}
  \country{Singapore}
}
\email{davidlo@smu.edu.sg}

\ccsdesc[500]{Software and its engineering~Search-based software engineering}

\keywords{Hybrid Code Retrieval, Cross-Language, Modality Collaboration, Contrastive Learning}

\begin{abstract}
  Effective code retrieval is indispensable and it has become an important paradigm to search code in hybrid mode using both natural language and code snippets. 
  Nevertheless, it remains unclear whether existing approaches can effectively leverage such hybrid queries, particularly in cross-language contexts.
  We conduct a comprehensive empirical study of representative code models and reveal three challenges: \textbf{(1)} insufficient semantic understanding; \textbf{(2)} inefficient fusion in hybrid code retrieval; and \textbf{(3)} weak generalization in cross-language scenarios.
  To address these challenges, we propose \textbf{\Ours}, a novel self-supervised framework designed to learn \textbf{Uni}fied \textbf{Co}de \textbf{R}epresentations that are semantically robust, modally collaborative, and language-agnostic.
  Firstly, we design a multi-perspective supervised contrastive learning module to enhance semantic understanding and modality fusion.
  It aligns representations from multiple perspectives, including code-to-code, natural language-to-code, and natural language-to-natural language, enforcing the model to capture a semantic essence among modalities.
  Secondly, we introduce a representation distribution consistency learning module to improve cross-language generalization, which explicitly aligns the feature distributions of different programming languages, enabling language-agnostic representation learning. 
  Extensive experiments on both an empirical benchmark and a large-scale benchmark show that \Ours outperforms all baseline models, achieving an average improvement of 8.64\% in MRR and 11.54\% in MAP over the best-performing baseline.
  Furthermore, \Ours exhibits stability in hybrid code retrieval and generalization capability in cross-language scenarios.
\end{abstract}

\maketitle

\input{Introduction}
\input{Emprical_Study}
\input{Approach}
\input{Experiment}
\input{RelationWork}
\input{Discussion}

\bibliographystyle{ACM-Reference-Format}
\bibliography{ref}

\end{document}

%% file: Introduction.tex
\section{Introduction}

With the increasing complexity of modern software and the thriving open source community, software codebases are expanding at an unprecedented rate.
Code retrieval, the technique of retrieving relevant code snippets from a codebase based on a given query, has become indispensable during the software development lifecycle \cite{ACMCS2023Kim-CodeSearchSurvey, TOSEM2023Xie-SurveyCS}.
For example, it is essential for critical activities such as program comprehension, code reuse, vulnerability detection, and automated code generation \cite{FSE2025Wang-ragincodegeneration, CodeSearch2025FSE}. 

Existing code retrieval research has mainly focused on single-modal queries, 
e.g., search code using natural language \cite{Shi2023CoCoSoDA,li2025scodesearcher, AAAI2025LIU-CodeSearch} or code \cite{li2023zc3, FSE2025CodeCloneSurvey}.
In practice, developers would expect the search engine to return better results if queried with more content, and, recognized by prior studies, developers would search using natural language and code at the same time (i.e., \textbf{Hybrid Code Retrieval}) \cite{CodeSearch2025FSE, ACL2025Li-CoIR, CHI2009Brandt-hybrid, liu2021characterizing, ACMCS2023Kim-CodeSearchSurvey}.
Indeed, Brandt et al. \cite{CHI2009Brandt-hybrid} observed that queries that mix code and natural language increase from 14\% at the beginning of a search session to 42\% by the eighth query reformulation.
Many modern code search engines, such as Searchcode\footnote{https://searchcode.com/}, are capable of accepting both natural language and code as input through their main interfaces \cite{ACMCS2023Kim-CodeSearchSurvey}.
For example, developers would search for "calculate the factorial of a number". However, the search engine would return a large number of results\footnote{https://stackoverflow.com/search?q=calculate+the+factorial+of+a+number}, and developers need to reformulate the query by adding more content to specialize the query.
Then, if the developer reformulates the query into "calculate the factorial of a number for loop", the search results still cannot satisfy the developer's need.\footnote{https://stackoverflow.com/search?q=calculate+the+factorial+of+a+number+for+loop}
If the developer reformulates the query into the special use case of for loop, i.e., "calculate the factorial of a number for(;i<=number;i++)", the search engine would return nothing\footnote{\url{https://stackoverflow.com/search?q=calculate+the+factorial+of+a+number+for\%28\%3Bi\%3C\%3Dnumber\%3Bi\%2B\%2B\%29}}, indicating that the search engine of Stack Overflow cannot understand the queries with code and natural language.

Yet, it remains an open question whether existing approaches can effectively utilize such hybrid queries. 
The increasingly popular multi-language software development further exacerbates this challenge.
In this architecture, developers need to constantly verify functionally equivalent code to ensure logical consistency across the entire system \cite{Nafi2019CLCDSACL, ICSE2024Chen-CCB}.
Although recent work such as CodeBridge \cite{liang2025codebridge} has revealed the potential complementarity between modalities and made progress by fusing the matching scores of multiple pre-trained models in cross-domain code search (such as SQL or Solidity), 
we still lack a systematic understanding of existing approaches to hybrid code retrieval, particularly in the more challenging cross-language scenario.

To address this critical gap, we first conduct a comprehensive empirical study that systematically evaluates the performance of seven representative pretrained code models or domain-specific state-of-the-art models under single-modal and hybrid code retrieval scenarios.
Our study aims to answer the following research questions:
(1) \textbf{RQ1: How do existing code models perform under different retrieval strategies?}
Here, we aim to investigate whether hybrid code retrieval can yield better results.
(2) \textbf{RQ2: How do existing code models perform in cross-language scenarios? }
Here, we aim to determine whether these models only learn the patterns that are specific to the language.
Our findings pinpoint three fundamental and interrelated challenges confronting the hybrid code retrieval:

\begin{enumerate}[labelsep=0.5em]
    \item \textbf{Insufficient Semantic Understanding}.
    Existing code models may excel at finding the first correct answer (measured by MRR) but struggle to appropriately find all relevant results (measured by MAP).
    One possible reason is that the existing models are dominated by surface lexical co-occurrences instead of semantic understanding.
    A better semantic understanding would enable the model to establish a better mapping between natural language and code.

   \item \textbf{Ineffective Fusion in Hybrid Code Retrieval}.
    We observe that the incorporation of natural language descriptions within code queries results in an average enhancement of 0.13\% in terms of MAP, which is considered to be marginal.
    This may be because existing code models cannot achieve modality complementarity in hybrid code retrieval.
    The true potential of hybrid code retrieval can only be realized when the model is able to perform modality fusion.

    \item \textbf{Weak Generalization in Cross-Language Scenarios}. 
    We observe that the average MRR for Code2Code retrieval declines from 64.2\% in the intra-language scenario (Python-to-Python) to 49.82\% in the cross-language scenario (Python-to-Java).
    This shows that the representations learned by the model are not language-agnostic. 
    Although hybrid retrieval can improve performance, it does not solve this problem. 
    If the model has language-agnostic representations, it will help improve the performance in cross-language scenarios.
\end{enumerate}

Motivated by these empirical findings, this paper introduces \Ours, a novel self-supervised framework designed to learn unified code representations that are semantically robust, modally collaborative, and language-agnostic. The framework is composed of two innovative modules:
Firstly, to overcome the challenges of insufficient semantic understanding and ineffective fusion, we design a \textbf{multi-perspective supervised contrastive learning module}. By constructing functionally equivalent yet formally diverse positive samples and optimizing across three perspectives: within code (Code2Code), within natural language (NL2NL), and cross-modal (NL2Code), this process forces the model to learn unified latent semantics, achieving deep modality alignment and collaboration.
Secondly, to tackle the challenge of weak generalization in cross-language scenarios, we introduce a \textbf{representation distribution consistency learning module}.
Inspired by domain adaptation theory \cite{ICSE2022CDCS}, this module employs Maximum Mean Discrepancy to align the representation spaces of different programming languages. 
It explicitly constrains functionally similar code, regardless of language, to share a consistent statistical distribution in the representation space. This guides the model to capture the logical essence of code rather than its language form, thereby enhancing cross-language generalization.

Extensive experiments on a comprehensive benchmark spanning standard empirical datasets and eleven programming languages validate our approach. The results demonstrate that \Ours not only significantly outperforms all baseline models but, more importantly, also exhibits superior stability in hybrid code retrieval and stronger generalization in cross-language scenarios, thereby effectively mitigating the core challenges identified in our empirical study.

In general, the contributions of this paper are as follows:
\begin{itemize}
    \item We conduct the first comprehensive empirical study that systematically reveals the inherent limitations of current hybrid code retrieval techniques in terms of semantic understanding, modality interaction, and cross-language generalization.
    
    \item We propose a novel self-supervised learning framework, named \Ours, which introduces a multi-perspective supervised contrastive learning mechanism and a representation distribution consistency learning module. 
    These innovations offer a new approach for learning unified and robust code representations.
    
    \item We validate the effectiveness of \Ours on a large-scale, multilingual benchmark. 
    The results not only demonstrate state-of-the-art performance across various retrieval tasks but also provide new perspectives and strong baselines for future research in code representation learning.
\end{itemize}

The remainder of this paper is organized as follows. Section \ref{sec:empirical} presents the empirical study. Section \ref{sec:method} overviews our proposed approach. The experimental setup and results are then described in Sections \ref{sec:EvaluationSettings} and \ref{sec:Evaluation}, respectively.  We present the relevant work in Section \ref{sec:relatedwork} and conclude in Section \ref{sec:conclusion}.

%% file: Emprical_Study.tex
\section{Empirical Study}
\label{sec:empirical}
To systematically investigate the challenges of existing code models in the hybrid code retrieval task, we conducted a comprehensive empirical study.

\subsection{Datasets}
\label{NL-Generation}
Effective evaluation of hybrid code retrieval relies on a benchmark containing functionally equivalent code pairs and corresponding high-quality, unique natural language descriptions. 
However, existing public datasets rarely provide these associations in a standardized format.
Therefore, we construct an empirical benchmark by integrating four widely-used datasets, i.e., CodeJamData \cite{Nafi2019CLCDSACL}, AtCoder \cite{Nafi2019CLCDSACL}, XLCoST \cite{Zhu2022XLCoSTAB}, and CodeSearchNet \cite{guo2022unixcoder}. 
The first two are derived from programming contests, and each problem description (NL) is mapped to multiple functionally equivalent code submissions (Code), establishing NL-Code and Code-Code relationships.
The latter two datasets contain fine-grained parallel data collected from open-source projects, such as different language implementations of the same function and their corresponding comments.

Although the above benchmark provides the necessary relationships, the quality of their native natural language descriptions is inconsistent. 
Comments from open-source projects are often too simple (e.g., "array sorting" and "two pointer technology"), which makes it difficult to accurately reflect the core functional intent of the code.
As a result, it creates an ineffective evaluation that rewards superficial topics rather than semantic understanding of code.
To solve this problem, we follow prior studies \cite{Internetware2023MCodeSearcher, liang2025codebridge, TOSEM2023RAPID, AAAI2024AdaCCD} and leverage an advanced code language model (e.g., Qwen2.5-Coder-7B-Instruct \cite{Hui2024Qwen25CoderTR}) to automatically generate concise, high-quality natural language queries for each code snippet as its standardized functional description.
The prompt template used for generation is as follows:\texttt{"Below is a \{lang\} code. Please give a short and accurate query the purpose of the code. You must write only summary without any prefix or suffix explanations. \{code\}"}. 
This process yielded an empirical benchmark with a unified format, where each sample consists of a code snippet and its standard natural language description, providing support for our retrieval evaluations.

\subsection{Baselines}
To ensure a comprehensive evaluation, seven representative baseline models were selected, covering both general pre-trained code models and domain-SOTA baseline models.

\textbf{General Pre-trained Code Models} include: \textbf{CodeBERT} \cite{feng2020codebert}, a bimodal pre-trained model for code and natural language via masked language modeling and replaced token detection tasks; \textbf{GraphCodeBERT} \cite{guo2020graphcodebert}, which enhances CodeBERT by incorporating data flows to understand code structure better; 
and \textbf{UniXcoder} \cite{guo2022unixcoder}, a unified cross-modal model that integrates Abstract Syntax Trees to support various code-related tasks.

\textbf{Domain-SOTA Baseline Models} include: 
For natural language to natural language retrieval, we use \textbf{BGE} \cite{bgeembedding}, a high-quality text embedding model trained via contrastive learning, specializing in accurately capturing semantic similarity. 
For natural language to code retrieval, we employ \textbf{CoCoSoDa} \cite{Shi2023CoCoSoDA}, which significantly improves the performance of the NL2Code task via the Momentum Contrastive Learning mechanism and data augmentation strategies.
For code-to-code retrieval, we utilize \textbf{ZC3} \cite{li2023zc3}, which builds on CodeBERT and achieves superior cross-lingual code representation via domain-aware learning and cycle consistency learning.
For hybrid code retrieval, we select \textbf{CodeBridge} \cite{liang2025codebridge}, to the best of our knowledge, which is the only publicly available method that can utilize both natural language to code and code to code. 
For a fair comparison, we use its fusion mechanism without LLM-based data augmentation, as the empirical benchmark already contains the corresponding data.

\begin{table*}[h]
\caption{
Code Retrieval Performance under Single Code Retrieval (left) and Hybrid Code Retrieval (middle) scenarios. The right part shows the optimal weights in the Weighted Fusion strategy. All values are percentages (\%). 
N and C refer to NL and Code, respectively.
From here on, the max value in each row is highlighted in bold, while the minimum is underlined.
}
\vspace{-0.3cm}
\label{tab:retrieval_results}
\small
\begin{tabular}{c  rr rr rr rr rr rr rr}
\toprule
\multirow{2}{*}{\textbf{Model}} & \multicolumn{2}{c}{\textbf{N2C}} & \multicolumn{2}{c}{\textbf{N2N}} & \multicolumn{2}{c}{\textbf{C2C}} & \multicolumn{2}{c}{\textbf{Remix}} & \multicolumn{2}{c}{\textbf{Concat}} & \multicolumn{2}{c}{\textbf{Weight}} & \multirow{2}{*}{\textbf{Weight-N2C}} &\multirow{2}{*}{\textbf{Weight-C2C}}\\
\cmidrule(lr){2-3} \cmidrule(lr){4-5} \cmidrule(lr){6-7} \cmidrule(lr){8-9} \cmidrule(lr){10-11} \cmidrule(lr){12-13}
& \textbf{MRR} & \textbf{MAP} & \textbf{MRR} & \textbf{MAP} & \textbf{MRR} & \textbf{MAP} & \textbf{MRR} & \textbf{MAP} & \textbf{MRR} & \textbf{MAP} & \textbf{MRR} & \textbf{MAP}\\
\midrule

CodeBert&\underline{7.80}&\underline{3.64}&\textbf{53.56}&\textbf{24.86}&24.88&10.41&20.33&8.52&15.55&6.84&24.97&10.44& 8.75($\pm$ 1.4) & 91.25($\pm$ 1.4)\\
GraphCodeBERT&\underline{28.84}&\underline{14.74}&\textbf{74.91}&\textbf{49.06}&50.73&29.19&51.50&29.09&50.98&28.89&53.57&30.58& 33.75($\pm$ 1.2) & 66.25($\pm$ 1.2)\\
UniXcoder&\underline{58.82}&\underline{36.98}&\textbf{84.74}&\textbf{64.20}&68.93&46.70&70.88&47.72&74.37&50.29&75.38&50.59& 46.25($\pm$ 8.1) & 53.75($\pm$ 8.1)
\\
BGE&\underline{17.08}&\underline{6.98}&\textbf{66.79}&\textbf{40.07}&24.98&12.25&24.56&11.14&25.36&11.58&26.87&12.72& 36.25($\pm$ 6.6) & 63.75($\pm$ 5.6)
\\
ZC3&\underline{49.03}&\underline{36.66}&77.74&57.31&83.97&69.00&\textbf{86.45}&\textbf{71.27}&83.59&68.31&85.67&70.75& 35.00($\pm$ 0.8) & 65.00($\pm$ 0.8)
\\
CoCoSoDa&\underline{46.97}&\underline{28.63}&\textbf{79.80}&\textbf{54.41}&55.77&36.68&57.81&37.27&60.45&39.02&61.48&39.44& 47.50($\pm$ 5.1) & 52.50($\pm$ 5.1)
\\

avg.&34.76&21.27&72.92&48.32&51.54&34.04&51.92&34.17&51.72&34.16&54.65&35.75& 34.58($\pm$ 3.9) & 65.42($\pm$ 3.9)\\
\midrule
CodeBridge&-&-&-&-&-&-&-&-&\underline{77.38} & \underline{53.83}&\textbf{81.21} & \textbf{56.38}& 20.00($\pm$ 1.2) & 31.25($\pm$ 12.9)\\
\bottomrule
\end{tabular}
\vspace{-0.1cm}
\end{table*}

\subsection{Code Retrieval: Definition and Strategies}
\label{problem}
Code retrieval is a representation learning-based ranking problem.
Given a query $q$ (a natural language description $NL$, a code snippet ${Code}$, or a combination), the goal is to retrieve the most semantically relevant code snippet $c_i$ from a large codebase $C=\{ c_1, c_2, \dots,c_N\}$.
An encoder $f(\cdot)$ is used to map both the query $q$ and candidate code snippet $c$ into a shared vector space $\mathbb{R}^d$, where semantic similarity is measured using cosine similarity:
\begin{equation}
    sim(q, c) = \frac{f(q) \cdot f(c)}{ || f(q) || \cdot || f(c) ||}
\end{equation}
Candidates in the codebase $C$ are then ranked based on their similarity scores, and the top-scoring snippet is returned.

This study focuses on evaluating hybrid code retrieval that uses natural language descriptions $Q_{nl}$ and code snippets $Q_{code}$ to jointly retrieve functionally similar codes. 
In addition to the three single-modal retrieval strategies of natural language to code (NL2Code), natural language to natural language (NL2NL), and code to code (Code2Code), given a code  $Q_{code}$ and its natural language description $Q_{nl}$, we also assess three hybrid approaches:
\begin{itemize}[wide=0pt]
\item \textbf{Input Remix}: Fuses queries at the input level by concatenating $Q_{nl}$ and $Q_{code}$ into a single hybrid input $Q_{Remix}$, which is then encoded to generate a query embedding.

\item \textbf{Vector Concat}: Fuses queries at the representation level. $Q_{nl}$ and $Q_{code}$ are independently encoded to respective vector representations, $V_{Q_{nl}}$ and $V_{Q_{code}}$, which are then concatenated into an augmented query vector $V_{concat-Q}=[V_{Q_{nl}};V_{Q_{code}}]$,
Correspondingly, the target code vector $V_{T_{code}}$ is self-concatenated to match query dimension $V_{concat-T}=[V_{T_{code}};V_{T_{code}}]$.

\item \textbf{Score Weight}: Inspired by CodeBridge\cite{liang2025codebridge}, this strategy fuses similarity scores.  
The hybrid similarity score $S$ is a weighted linear sum of the NL2Code and Code2Code similarities:
\begin{equation}
\label{eq1}
    S = \alpha S_{nl2code} + (1- \alpha) S_{code2code}
\end{equation}
The weighting coefficient $\alpha \in [0, 1]$ is selected via grid search to achieve optimal performance.

\end{itemize}

\subsection{Evaluation Metrics}
All experiments are evaluated using two standard code retrieval metrics: Mean Reciprocal Rank (MRR) and Mean Average Precision (MAP), which are commonly used in prior studies \cite{guo2022unixcoder, li2023zc3, Shi2023CoCoSoDA}.
MRR measures the average reciprocal rank of the first relevant result,
reflecting the capability of the model to retrieve the correct answer at the top of the list.
MAP provides a more comprehensive assessment by evaluating the ranking quality of all relevant results.
\subsection{Research Questions}
Our empirical study focuses on the following research questions:

\noindent\textbf{RQ1: How do existing code models perform under different retrieval strategies?}
We systematically evaluate baseline models in both single (NL2Code, NL2NL, Code2Code) and hybrid settings to understand whether searching using natural language and code can have a better result.

\noindent\textbf{RQ2: How do existing code models perform in cross-language scenarios?}
We evaluate model performance in both intra-language (e.g., Python-to-Python) and cross-language (e.g., Python-to-Java) retrieval scenarios to measure their capability to generalize across different programming languages.

\subsection{Empirical Findings}
\subsubsection{RQ1: How do existing code models perform under different retrieval strategies?}
Table \ref{tab:retrieval_results} presents the performance of seven baseline models across different retrieval strategies. The results reveal two key insights:

Firstly, \textbf{NL2Code is less effective than Code2Code or NL2NL in single strategies}. 
As shown in Table \ref{tab:retrieval_results}, 
different models excel at specific retrieval strategies. 
For instance, ZC3 demonstrates superior code representation capabilities, achieving the highest MRR of 83.97\%, which indicates robust code representation capabilities.
Conversely, UniXcoder leads in both NL2NL (MRR 84.74\%) and NL2Code (MRR 58.82\%) strategies.
Notably, most code models, except CodeBERT, outperform the specialized text embedding model BGE (MRR 66.79\%) on the NL2NL strategy.
This suggests that the \textbf{pre-training on the logical structure of code can enhance their understanding of natural language descriptions related to code}.
Crucially, as underlined in Table 1, we observe that all models perform poorly on the NL2Code strategy.
This emphasises the difficulties that current models encounter when attempting to bridge the semantic gap between natural language and code.

Secondly, contrary to intuition, \textbf{hybrid code retrieval cannot bring additional benefits compared with single retrieval, with performance being highly dependent on model and strategy.}
Although the Remix strategy incorporates code information into the query input, its MAP score improves by only 0.13\% compared to the Code2Code baseline.
This demonstrates the difficulty of \textbf{existing models in effectively utilising the complementary information between natural language and code}.
The Weight strategy is the only one that can stably improve the performance across all models, but the improvement is quite limited (an average improvement of MRR is 3.1\%, and an average improvement of MAP is 1.7\%).
Moreover, the variance of the optimal weight $\alpha$ across different datasets averages around 3.9\%, indicating that the strategy lacks generalizability and necessitates task-specific hyperparameter tuning.
It is worth noting that ZC3 is an exception, achieving the best hybrid code retrieval performance (MRR 86.45\%) with the Remix strategy. 
This can be attributed to its extremely powerful code understanding capability, which enables the additional natural language to serve as an effective supplement. 
However, even CodeBridge, which integrates multiple SOTA models (BGE, UniXCoder, CoCoSoDA), still underperforms ZC3.
This finding shows that \textbf{simply ensembling powerful models is insufficient to overcome the fundamental challenges of hybrid code retrieval}.

\begin{center}
    \vspace{-0.2cm}
    \fcolorbox{black}{green!5}{\parbox{0.98\linewidth}{
    \textbf{Finding 1: Ineffective Fusion in Hybrid Code Retrieval}.
    Existing code models have specific strengths: ZC3 performs best in Code2Code retrieval (MRR 83.97\%), and UniXcoder performs best in NL2NL retrieval (MRR 84.74\%).
    However, the Weight strategy achieves only a modest average MRR improvement of 3.1\% and is not robust, as the optimal weight $\alpha$ varies by more than 3.9\% across datasets. These results suggest that current fusion methods are unable to leverage the complementary information from both modalities effectively.
    }}
\end{center}

\begin{table}[h!]
\centering
\vspace{-0.3cm}
\caption{Performance in cross-language scenarios.
}
\vspace{-0.3cm}
\label{tab:retrieval_cross}
\small
\resizebox{\columnwidth}{!}{
\begin{tabular}{c rr rr rr rr}
\toprule
\multirow{2}{*}{\textbf{ Strategy }} & \multicolumn{2}{c}{\textbf{Java2Java}} & \multicolumn{2}{c}{\textbf{Python2Python}} & \multicolumn{2}{c}{\textbf{Java2Python}} & \multicolumn{2}{c}{\textbf{Python2Java}}\\
\cmidrule(lr){2-3} \cmidrule(lr){4-5} \cmidrule(lr){6-7} \cmidrule(lr){8-9}
& \textbf{MRR} & \textbf{MAP} & \textbf{MRR} & \textbf{MAP} & \textbf{MRR} & \textbf{MAP} & \textbf{MRR} & \textbf{MAP} \\
\midrule
Code2Code&46.59&21.11&\textbf{64.20}&34.59&\underline{41.13}&29.98&49.82&32.26\\
Remix&47.80&22.51&\textbf{59.49}&32.68&\underline{46.06}&32.26&50.27&32.13
\\
Concat&46.91&22.20&\textbf{61.00}&33.70&\underline{45.52}&32.12&49.87&31.99\\
Weight&51.35&23.21&\textbf{66.09}&35.73&\underline{46.78}&33.28&51.32&33.09
\\
\bottomrule

\end{tabular}
}
\vspace{-0.3cm}
\end{table}

\subsubsection{RQ2: How do existing code models perform in cross-language scenarios?}
As illustrated in Table \ref{tab:retrieval_cross}, the model performance is evaluated in both intra-language and cross-language scenarios. 
Detailed results for all models are available in our published artifacts.
The results reveal that:

Firstly, \textbf{all evaluated models demonstrate inadequate generalization under cross-language code retrieval, and the hybrid code retrieval has not effectively alleviated this shortcoming}.
For instance, the average MRR for Code2Code retrieval declines from 64.2\% in the intra-language scenario (Python-to-Python) to 49.82\% in the cross-language scenario (Python-to-Java).
This difference in performance reflects that model representations are closely related to the specific language.
More importantly, \textbf{while the Weight strategy improves performance, it also exacerbates performance imbalances in different scenarios.}
As shown in Table \ref{tab:retrieval_cross}, the implementation of this strategy results in a 1.89\% MRR improvement for the intra-language scenario (Python-to-Python) in comparison to the Code2Code baseline, while the improvement for the cross-language scenario (Python-to-Java) is marginal at 1.5\%.
\begin{center}
\vspace{-0.2cm}
    \fcolorbox{black}{green!5}{\parbox{0.98\linewidth}{
    \textbf{Finding 2: Weak Generalization in Cross-Language Scenarios}: 
    Existing models demonstrate poor generalization in cross-language retrieval.
    Hybrid strategies fail to resolve this issue and can even exacerbate performance imbalances in different scenarios. 
    This suggests that models only learn language-specific representations, and developing language-agnostic representations remains an unsolved challenge.
    }}
\end{center}

Secondly, a contradictory trend is observed: \textbf{MRR is inconsistent with MAP}. 
For example, when switching from cross-language to intra-language retrieval, the MRR of the Code2Code strategy with Java queries increases by 5.46\%, whereas MAP declines by 8.87\%.
This divergence is significant because MRR measures the rank of the first correct result, whereas MAP, which evaluates the ranking of all relevant results, serves as a proxy for better semantic comprehension. 
This trend suggests that when the query and target languages are the same, models can easily identify one correct answer via surface-lexical matching, thus increasing MRR scores.
However, this reliance on surface lexical prevents the model from understanding the deeper functional semantics required to identify all other related but syntactically distinct code snippets, thus reducing MAP scores.
This finding suggests an imbalance in the learned representations, characterised by a reliance on surface lexical features relative to deep logical understanding.
 
\begin{center}
\vspace{-0.3cm}
    \fcolorbox{black}{green!5}{\parbox{0.98\linewidth}{
    \textbf{Finding 3: Insufficient semantic understanding}. 
    The model representations evaluated are susceptible to surface lexical co-occurrences while demonstrating a limited understanding of the underlying equivalent functional intent.
    This results in a misalignment of the semantic representations, thereby exposing the model's limitations in comprehending semantic equivalence at the functional level.
    }}
\end{center}

%% file: Approach.tex
\begin{figure*}[h]
    \centering
    \includegraphics[width=0.9\textwidth]{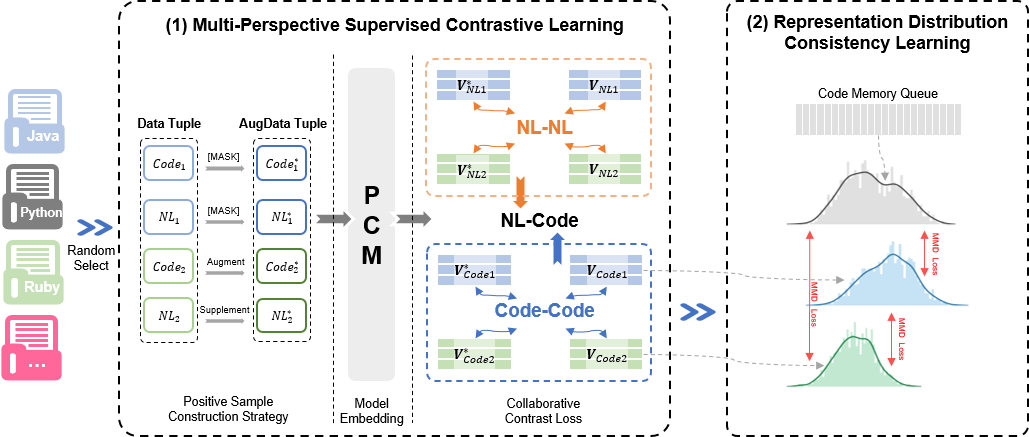}
    \caption{The overall framework of \Ours. \Ours first randomly selects a language pair, then uses various data augmentation methods to 
    construct diverse positive sample pairs. 
    MPCL is designed to learn a feature representation in which modal features interact fully. Finally, RDCL is designed to reduce differences between the features of different languages.}
    \label{fig:FrameWork}
    \vspace{-0.1cm}
\end{figure*}
\section{Approach}
\label{sec:method}
As previous empirical analyses have revealed, existing code representation models face three challenges, i.e., insufficient semantic understanding, ineffective fusion in hybrid code retrieval, and weak generalization in cross-language scenarios. 

To address these challenges, we propose a novel self-supervised code representation learning framework called \Ours. 
This framework aims to enhance hybrid code retrieval performance by learning a semantically robust, modally collaborative, and language-agnostic representation through end-to-end joint optimization.
The core architecture of \Ours (shown in Figure \ref{fig:FrameWork}) consists of two innovative modules:

\begin{itemize}[wide=0pt]
    \item \textbf{Multi-Perspective Supervised Contrastive Learning.} To address the challenges of \textbf{insufficient semantic understanding} and \textbf{ineffective fusion}, 
    this module is designed to reduce the co-occurrence probability of sample words and phrases, forcing the model to go beyond superficial word co-occurrence and learn the deep computational logic behind the code by constructing functionally equivalent but formally diverse positive samples. 
    It conducts joint contrastive learning from three perspectives, i.e., code ( Code2Code ), natural language (NL2NL), and cross-modal (NL2Code),
    to build a unified representation space with modality alignment.

    \item \textbf{Representation Distribution Consistency Learning.} 
    To address the challenge of weak generalization in cross-language scenarios, 
    this module employs the Maximum Mean Discrepancy (MMD) method \cite{TMLR2012MMD} to align the representation distributions of different programming languages, drawing on domain adaptation theory.
    This encourages the model to focus on the logical essence of the code rather than its grammatical form.
\end{itemize}

This section details the components of the \Ours framework and their underlying mechanisms.

\subsection{Multi-Perspective Supervised Contrastive Learning}
\label{subsec:MPCL}
To address the limitations of pretrained code models (PCMs) in semantic understanding and the lack of cross-modal interaction in hybrid code retrieval, we design a multi-perspective supervised contrastive learning module. 
The core idea is to achieve deep semantic understanding and modality alignment within a unified representation space by leveraging carefully constructed positive sample sets and a collaboratively optimized multi-perspective loss function. 
This module is built upon two key components: a multi-source positive sample construction strategy and a multi-perspective collaborative contrast loss.

\subsubsection{Multi-Source Positive Sample Construction Strategy}
High-quality positive pairs are essential for contrastive learning.
To force the model to go beyond superficial co-occurrence of words and learn deep functional semantics, we construct a rich and diverse set of positive samples from the following three levels:
\begin{enumerate}[wide=0pt]
    \item \textbf{Functionally equivalent data sources}. Functionally similar data tuples $D=(Code_1, Code_2, NL_1, NL_2)$ are randomly sampled from different programming languages, where $Code_1$ and $Code_2$ are functionally equivalent but different implementations of code snippets, and $NL_1$ and $NL_2$ are their corresponding descriptions. This naturally forces the model to ignore specific syntax or keywords, thereby enhancing the robustness of semantic understanding.
    \item \textbf{Data augmentation and perturbation}. To guide the model in learning fine-grained semantic invariance, inspired by CoCoSoDA \cite{Shi2023CoCoSoDA}, we perform random data augmentation (e.g., dynamic masking, identifier replacement, etc.) on data tuples $D$ to generate semantically consistent but formally slightly different perturbed positive samples $D^*$.
    \item \textbf{Cross-modal positive sample supplementation}.  
    To address the limited natural language enhancement methods, which lead to insufficient diversity of positive samples of NL,
    code comments are also treated as corresponding descriptions to enrich the semantic expression.
\end{enumerate}

Through the above strategy, a rich set of positive samples is constructed for each element in the data tuple $D$.
For example, the positive code sample set for $Code_1$ is $\{Code_1^*, Code_2, Code_2^*\}$, and the positive natural language sample set for $Code_1$  is $\{NL_1,NL_1^*, NL_2, NL_2^*\}$.

\subsubsection{Multi-Perspective Collaborative Contrast Loss}
Building on diverse positive samples, three complementary contrastive learning objectives are designed to promote deep interaction between modalities and lay a solid foundation for hybrid code retrieval.
The model is optimized within the Momentum Contrast (MoCo) framework \cite{He2019moco}, which maintains dynamic queues ($queue_{code}$, $queue_{code}^*$, $queue_{nl}$, $queue_{nl}^*$) for each modality to provide large-scale and consistent samples, thereby stabilizing the training process.

\begin{itemize}[wide=0pt]
    \item Code-to-Code Comparison: enforcing semantic invariance by encouraging the model to map functionally equivalent yet syntactically different code snippets (e.g., $Code_1$ and $Code_2$) closer in the representation space. This forces the model to ignore surface-level syntactic differences and focus on the underlying functional semantics.
    
    \item Natural Language-to-Natural Language Comparison: enhancing linguistic nuance understanding by bringing together heterogeneous natural language texts (e.g., $NL_1$ and $NL_2$) that describe the same functionality, encouraging the model to move beyond keyword matching and capture the equivalence in expressed intent.
    
    \item Natural Language-to-Code Comparison: building a cross-modal bridge by aligning natural language descriptions with their corresponding code implementations,
    establishing a robust cross-modal mapping between intent and functionality.

\end{itemize}

For each positive pair $(q, k^+)$, 
the InfoNCE loss is employed to maximize their similarity while simultaneously minimizing the similarity between $q$ and all negative samples $k^-$:
\begin{equation}
    L_{CL} = -log\frac{exp(\text{sim}(f(q), f(k^+)))}{exp(\text{sim}(f(q), f(k^+))) + \sum_{k^-}exp(\text{sim}(f(q), f(k^-)))}
\end{equation}
where $f(\cdot)$ denotes the backbone model, and $k^-$ are negative samples drawn from other instances in the current batch and the dynamic queue.

The multi-perspective supervised contrastive loss, $L_{MPCL}$, is the sum of the losses from the three contrastive objectives described above.
This guides the model to learn a unified representation that satisfies all perspective constraints. 
The model distinguishes between implementation and function in code (Code2Code), between linguistic expression and intent (NL2NL), and establishes a strong alignment between the two modalities (NL2Code).

\begin{equation}
    L_{MPCL} = L_{Code2Code-CL} + L_{NL2NL-CL}+ L_{NL2Code-CL}
\end{equation}

\subsection{Representation Distribution Consistency Learning}
\label{subsec:RDCL}

To address the limited generalization capabilities of models in cross-language tasks, we propose a representation distribution alignment module. 
This module aims to guide code from different programming languages to form similar feature distributions in the embedding space. 
Maximum Mean Discrepancy (MMD) \cite{TMLR2012MMD} is introduced as a non-parametric metric for distribution distance and is seamlessly integrated into the contrastive learning framework. 
To achieve comprehensive and robust distribution alignment, the MMD Loss is applied at both local and global levels for complementary alignment.

\begin{itemize}[wide=0pt]
    \item \textbf{Local Batch Distribution Alignment}. Drawing inspiration from domain transfer in meta-learning, each training batch is constructed to include code pairs from different programming languages, where $Code_1$ originates from language A and $Code_2$ from language B. Within each batch, the MMD loss is computed between the code sets $Code_1$ and $Code_2$. This approach encourages the model to generate language-agnostic representations in real time while processing each local batch of data.

    \begin{equation}
        L_{local} = \text{MMD}(f({Code_1}),f(Code_2))
    \end{equation} 

    \item  \textbf{Global Historical Distribution Alignment}. Local alignment alone may cause the representation space to drift across different training epochs. To ensure consistency of the global distribution, the MMD loss is further computed between the current batch’s code sets ($Code_1$ and $Code_2$) and the code dynamic queues $queue_{code}$. This queue serves as a stable estimate of the historical distribution over the entire dataset. 
    This loss term ensures that while learning features of specific language pairs, the local representation distribution remains aligned with the historical, thereby preventing distribution drift and catastrophic forgetting.

    \begin{equation}
    \begin{aligned}
            L_{global} = \text{MMD}(f({Code_1}),f(queue_{code})) + \\
    \text{MMD}(f({Code_2}),f(queue_{code}))
    \end{aligned}
    \end{equation}

\end{itemize}
The representation distribution consistency learning loss, $L_{RDCL}$, is the sum of the losses from these two levels.
\begin{equation}
    L_{RDCL} = L_{local} + L_{global}
\end{equation}

By minimizing $L_{RDCL}$, the encoder $f(\cdot)$ is explicitly constrained to disregard language-specific syntactic features and instead focus on the deeper, language-agnostic "semantic core" shared across all languages, thereby constructing a truly language-agnostic representation space.

\subsection{Overall training objectives}
The overall loss function of \Ours is the sum of the multi-perspective supervised contrastive loss and the representation distribution consistency learning loss.
\begin{equation}
    L = L_{MPCL} + L_{RDCL}
\end{equation}

By optimizing this objective function end-to-end, the framework learns a deeply integrated, semantically rich, and highly generalizable code representation, thereby systematically addressing the three aforementioned challenges. 
The momentum encoder and the dynamic queues are updated after each gradient step.

%% file: Experiment.tex
\section{Evaluation Settings}
\label{sec:EvaluationSettings}
To evaluate the effectiveness of the proposed framework, \Ours, 
we conducted a series of experiments.
This section presents the core research questions, the used datasets, the baseline models for comparison, the evaluation metrics, and the implementation details.

 \subsection{Research Questions}
To systematically evaluate \Ours, we consider the following two core research questions to validate its effectiveness.

\noindent\textbf{RQ3: What is the overall performance of \Ours?} 
Here, we evaluate the effectiveness and generalization of \Ours by comparing its performance against a set of baseline models on the empirical study benchmark as well as on a large-scale cross-language test dataset.

\noindent\textbf{RQ4: How effective are the key designs of \Ours, and in what aspects are its strengths demonstrated?}
Here, we conduct a series of experiments, including ablation studies, generalizability studies, modality co-analysis, parameter sensitivity analysis, and case studies, to explore the internal mechanisms behind the success of \Ours and its specific advantages in addressing core challenges.

\subsection{Datasets}
\textbf{Train data.} To ensure fairness, we followed previous work \cite{li2023zc3} and used the code translation dataset from CodeNet \cite{Puri2021CodeNet} for training. This dataset covers three programming languages (i.e., Ruby, Python, and Java) and includes a total of 50,868 function-level implementations addressing 4,053 distinct programming problems. We retained its original settings and used the data corresponding to the first 3,500 problems for model training.

\textbf{Test data}. To ensure a comprehensive and challenging evaluation, two sets of test datasets are employed: 
(1) The empirical study benchmark, which includes CodeJamData \cite{Nafi2019CLCDSACL}, AtCoder \cite{Nafi2019CLCDSACL}, XLCoST \cite{Zhu2022XLCoSTAB}, and CodeSearchNet \cite{guo2022unixcoder}. 
(2) To further evaluate the model's generalization capability in complex cross-language scenarios, we constructed a new dataset from the large-scale multilingual benchmark XCodeEval \cite{khan2024xcodeeval}. 
This test dataset carefully is selected from functionally similar code pairs across 11 mainstream programming languages, strategically covering: 
1) Languages encountered during backbone pretraining (e.g., Java, Python, Go, Ruby, JavaScript, PHP); 
2) Other frequently used languages (e.g., C, C++, C\#); 
and 3) Languages unseen by the model during training (e.g., Rust, Scala). 

All evaluation data underwent a rigorous screening process: 
1) \textbf{Static analysis filtering}. Tree-sitter is used to remove code samples with syntax errors or incomplete structures; 
2) \textbf{Mitigation of testing bias}. To avoid evaluation results being influenced by the frequent occurrence of common functions (such as quick sort), we randomly select code samples that implement similar functions in the same language to ensure that no more than 10 code samples are retained for each function;
3) \textbf{Data Deduplication for Unbiased Evaluation}. To ensure the validity of the evaluation, the train and test data were deduplicated to prevent overlap and thus avoid potential bias in the test results.

\begin{table}
\centering
\caption{Overview of the test dataset}
\label{tab:dataset-statistics}
\vspace{-0.3cm}
\resizebox{\columnwidth}{!}{
\begin{tabular}{cc cc cc}
\toprule
\textbf{Dataset} & \textbf{\#Languages} & \textbf{\#Datasize} & \textbf{\#Problems} & \textbf{Code Tokens} & \textbf{NL Tokens} \\
\midrule
CodeJamData & 2 & 402 & 21 & 764 & 75\\
AtCoder & 2 & 1386 & 77 & 517 & 62\\
XLCoST & 2 & 882 & 882 & 215 & 54\\
CodeSearchNet & 2 & 3148 & 324 & 708 & 60\\
XCodeEval & 11 & 20148 & 6574 & 274 & 69 \\
\bottomrule
\end{tabular}
}
\vspace{-0.5cm}
\end{table}

Table \ref{tab:dataset-statistics} presents key statistics for all evaluation datasets, where \textbf{\#Languages} denotes the total number of programming languages covered, \textbf{\#Datasize} denotes the average number of code–text sample pairs per language, \textbf{\#Problems} denotes the total number of functionally independent problems, and \textbf{Code Tokens} and \textbf{NL Tokens} denote the average token counts for code and natural language descriptions, respectively. This multidimensional and multilingual evaluation setup enables us to thoroughly examine the robustness of the model's semantic understanding and its generalization capabilities in cross-language scenarios.

\subsection{Baseline Models, Evaluation Metrics, and Implementation Details}
\textbf{Baseline Models}:
To ensure a comprehensive evaluation, baseline models cover all models we used in section \ref{sec:empirical}, including general pre-trained code models, domain-SOTA baselines models.
Building upon this foundation, we further introduce several methods as new baselines:
1) \textbf{Bag-of-Words-based Retrieval Method}. BM25 \cite{Robertson2009BM25} is an enhanced text-matching algorithm based on TF-IDF, serving as a strong baseline for unsupervised code search.
2) \textbf{Zero-shot LLM-based Methods}. This category includes SGPT \cite{patel2024evaluating}, which generates high-quality sentence-level embeddings by weighted averaging of token embeddings across LLM layers, and LLM2VEC \cite{COLM2024llmvec}, a technique that efficiently converts decoder-only LLMs into powerful bidirectional embedding models. We apply LLM2VEC to Deepseek-Coder-7b-instruct-v1.5 \cite{guo2024deepseekcoderlargelanguagemodel} and Qwen2.5-Coder-7B-Instruct \cite{Hui2024Qwen25CoderTR}, denoted as DS2Vec and Qwen2Vec, respectively.
3) \textbf{SOTA Commercial Embedding Models}. The text-embedding-3-large \cite{openai2024embedding}, released by OpenAI in 2024, provides 3072-dimensional dense representations optimized for semantic retrieval and cross-domain generalization. For convenience, we refer to it as TE3L, representing the current industrial SOTA in text and code retrieval.

\textbf{Evaluation metrics:} The performance of all models is evaluated using standard metrics in the code retrieval domain, i.e., Mean Reciprocal Rank (MRR) and Mean Average Precision (MAP).

\textbf{Implementation details}: All models are evaluated in a zero-shot setting, without any fine-tuning on the downstream datasets.
For our proposed framework, \Ours, we use UniXcoder \cite{guo2022unixcoder} as the backbone encoder, chosen due to its demonstrated solid foundational capabilities in multi-modal coding tasks.
For \Ours and other pre-trained code models, the maximum sequence lengths of code and natural language descriptions are truncated or padded to 256 and 128 tokens \cite{guo2022unixcoder}, respectively. 
In contrast, the evaluation of zero-shot LLM methods is not subject to this length restriction.

During training, we use the AdamW optimizer \cite{AdamW} with a learning rate of $2\times10^{-5}$.
The batch size is 40 per GPU. 
The hyperparameters of MoCo follow the original setting \cite{He2019moco}. 
For the MMD component, we adopt a multi-scale Gaussian kernel with a bandwidth combination of $\sigma=\{0.6, 1.2, 2.4\}$. 
All experiments were conducted on a server equipped with two Tesla V100 GPUs (each with 32 GB of VRAM).

\section{Evaluation Analysis}
\label{sec:Evaluation}
\subsection{RQ3: What is the overall performance of \Ours?}
To answer RQ3, we conducted a comprehensive comparison of \Ours against all baseline models on the empirical benchmarks and the large-scale, cross-language test dataset, XCodeEval.
Each table reports the average performance across all language scenarios, and please refer to our artifacts for the detailed values.

\begin{table}
\centering
\caption{Comparison of Code Retrieval Performance on the Empirical Benchmark.
From here on, the max value in each column is highlighted in bold, while the second max is underlined.}
\vspace{-0.3cm}
\label{tab:retrieval_final}
\resizebox{\columnwidth}{!}{
\begin{tabular}{l rr rr rr rr rr rr}
\toprule
\multirow{2}{*}{\textbf{Model}} & \multicolumn{2}{c}{\textbf{NL2Code}} & \multicolumn{2}{c}{\textbf{NL2NL}} & \multicolumn{2}{c}{\textbf{Code2Code}} & \multicolumn{2}{c}{\textbf{Remix}} & \multicolumn{2}{c}{\textbf{Concat}} & \multicolumn{2}{c}{\textbf{Weight}} \\
\cmidrule(lr){2-3} \cmidrule(lr){4-5} \cmidrule(lr){6-7} \cmidrule(lr){8-9} \cmidrule(lr){10-11} \cmidrule(lr){12-13}
& \textbf{MRR} & \textbf{MAP} & \textbf{MRR} & \textbf{MAP} & \textbf{MRR} & \textbf{MAP} & \textbf{MRR} & \textbf{MAP} & \textbf{MRR} & \textbf{MAP} & \textbf{MRR} & \textbf{MAP} \\
\midrule
CodeBERT&7.80&3.64&53.56&24.86&24.88&10.41&20.33&8.52&15.55&6.84&24.97&10.44
\\
GraphCodeBERT&28.84&14.74&74.91&49.06&50.73&29.19&51.50&29.09&50.98&28.89&53.57&30.58\\
UniXcoder&58.83&36.98&84.74&64.20&68.93&46.70&70.88&47.72&74.37&50.29&75.38&50.59\\
BGE&17.08&6.98&66.79&40.07&24.98&12.25&24.56&11.14&25.36&11.58&26.87&12.72
\\
ZC3&49.03&36.66&77.74&57.31&\underline{83.97}&\underline{69.00}&\underline{86.45}&\underline{71.27}&83.59&\underline{68.31}&85.67&\underline{70.75}
\\
CoCoSoDa&46.97&28.63&79.80&54.41&55.77&36.68&57.81&37.27&60.45&39.02&61.48&39.44
\\
CodeBridge&-&-&-&-&-&-&-&-&77.38 & 53.83&81.21 & 56.38\\
\midrule
BM25&25.43&16.11&82.12&58.58&54.83&39.67&54.35&39.97&-&-&-&-\\
TE3L&\textbf{82.51}&\underline{61.89}&\underline{87.63}&\underline{70.18}&78.55&58.53&83.33&62.82&\underline{85.43}&65.41&\underline{87.13}&66.80\\
SGPT&44.91&28.36&76.00&50.73&66.49&44.41&67.37&45.35&67.37&44.77&69.26&46.51
\\
DS2Vec&11.36&5.47&70.80&42.67&55.26&32.25&33.15&19.80&31.50&18.11&55.51&32.15
\\
Qwen2Vec&33.62&18.21&77.19&53.15&67.33&43.38&69.58&44.65&66.99&41.12&70.24&44.24
\\
\midrule
\Ours&\underline{81.81}&\textbf{70.24}&\textbf{89.01}&\textbf{78.88}&\textbf{89.12}&\textbf{78.75}&\textbf{92.26}&\textbf{81.55}&\textbf{90.81}&\textbf{80.20}&\textbf{92.09}&\textbf{80.84}
\\
\bottomrule

\end{tabular}
}
\vspace{-0.3cm}
\end{table}

Table \ref{tab:retrieval_final} illustrates the performance of all approaches on the empirical benchmarks. 
The results demonstrate that \textbf{\Ours consistently outperforms all baseline models across all six retrieval strategies (both single and hybrid), 
achieving relative improvements of 3.66\% in MRR and 9.84\% in MAP over the best-performing baseline.}
\textbf{Comparison with pre-trained code models}: Compared to UniXcoder, \Ours achieves a substantial improvement on the most challenging NL2Code, an increase in MRR from 58.83\% to 81.81\% (a relative improvement of 39.1\%). This highlights the effectiveness of our proposed framework in significantly enhancing pre-trained models for retrieval tasks.
\textbf{Comparison with domain-specific SOTA models}: Even when compared with SOTA methods such as ZC3, CoCoSoDa, and CodeBridge, \Ours demonstrates a clear performance advantage. 
Notably, \Ours outperforms ZC3, which is optimized specifically for the Code2Code task, by 5.15\% in terms of MRR and 9.72\% in MAP.
Furthermore, under the hybrid code retrieval like the Weight, \Ours outperforms CodeBridge by 10.88\% in terms of MRR and 24.46\% in MAP.
This demonstrates that the learned unified representation is superior in quality and collaboration to all specialized post-fusion or single-task optimization strategies.
\textbf{Comparison with Zero-Shot Large Language Models}: Both SGPT and Qwen2Vec - the latter of which is built on a more powerful backbone - perform significantly worse than \Ours across all evaluation metrics. 
This may be due to the autoregressive training method of general LLMs and the emphasis on natural language, making it difficult for them to capture the structured semantics representation of code.
This demonstrates that for code retrieval, \Ours holds a distinct advantage over the zero-shot approaches from general-purpose LLMs.
\textbf{Comparison with SOTA Commercial Embedding Models}: Compared to TE3L, \Ours has fewer parameters yet achieves comparable or superior performance across all retrieval settings.
This demonstrates that \Ours learns representations that more effectively capture the fine-grained semantic and structural relationships in code.

\begin{table}[h!]
\centering
\vspace{-0.1cm}
\caption{Comparison of Code Retrieval Performance on XCodeEval Benchmark}
\label{tab:retrieval_final_XCE}
\vspace{-0.3cm}
\resizebox{\columnwidth}{!}{
\begin{tabular}{l rr rr rr rr rr rr}
\toprule
\multirow{2}{*}{\textbf{Model}} & \multicolumn{2}{c}{\textbf{NL2Code}} & \multicolumn{2}{c}{\textbf{NL2NL}} & \multicolumn{2}{c}{\textbf{Code2Code}} & \multicolumn{2}{c}{\textbf{Remix}} & \multicolumn{2}{c}{\textbf{Concat}} & \multicolumn{2}{c}{\textbf{Weight}} \\
\cmidrule(lr){2-3} \cmidrule(lr){4-5} \cmidrule(lr){6-7} \cmidrule(lr){8-9} \cmidrule(lr){10-11} \cmidrule(lr){12-13}
& \textbf{MRR} & \textbf{MAP} & \textbf{MRR} & \textbf{MAP} & \textbf{MRR} & \textbf{MAP} & \textbf{MRR} & \textbf{MAP} & \textbf{MRR} & \textbf{MAP} & \textbf{MRR} & \textbf{MAP} \\
\midrule
CodeBert&0.42&0.17&0.76&0.27&0.51&0.20&0.48&0.19&0.46&0.18&0.51&0.20\\
GraphCodeBERT&2.41&0.82&11.5&4.65&1.92&0.69&2.07&0.72&2.17&0.76&2.41&0.82\\
UniXcoder&15.88&7.51&48.57&26.72&17.67&8.31&18.75&8.93&23.28&11.19&23.28&11.19\\
BGE&1.55&0.44&24.10&9.36&2.25&0.71&2.17&0.62&2.27&0.69&2.36&0.73
\\
ZC3&24.40&15.84&41.45&22.95&\underline{53.98}&\underline{36.77}&\underline{57.10}&\underline{39.17}&\underline{52.21}&\underline{35.01}&\underline{56.38}&\underline{38.59}
\\
CoCoSoDa&6.94&2.68&32.69&15.41&7.04&2.68&8.15&3.14&9.58&3.78&9.58&3.78
\\
CodeBridge&-&-&-&-&-&-&-&-& 27.43& 13.14& 31.75& 14.82\\
\midrule
BM25& 1.80& 0.92& 44.05& 22.15& 7.60& 3.34& 6.94& 3.16&-&-&-&-\\
TE3L& \underline{41.79}& \underline{24.24}& \underline{57.16}& \underline{35.22}& 42.85& 24.03& 47.83& 27.71& 50.49& 29.78&50.49& 29.78\\
SGPT&9.66&3.84&33.71&15.74&12.87&5.36&16.20&6.85&15.60&6.52&15.76&6.49
\\
DS2Vec&0.86&0.34&36.21&16.58&10.44&4.06&4.07&1.49&3.42&1.19&10.31&3.78\\
Qwen2Vec&5.56&2.23&40.51&20.10&12.04&4.89&14.60&6.06&12.54&5.12&13.38&5.29
\\
\midrule
\Ours&\textbf{57.76}&\textbf{40.92}&\textbf{65.00}&\textbf{45.31}&\textbf{66.53}&\textbf{48.17}&\textbf{69.06}&\textbf{50.45}&\textbf{70.67}&\textbf{51.48}&\textbf{71.31}&\textbf{52.12}
\\
\bottomrule

\end{tabular}
}
\vspace{-0.2cm}
\end{table}

Table \ref{tab:retrieval_final_XCE} shows the results of all models on the more challenging cross-language benchmark, XCodeEval. \textbf{The results show that \Ours exhibits excellent generalization and robustness,
achieving relative improvements of 13.62\% in MRR and 13.24\% in MAP over the best-performing baseline.}
As shown in the table, the performance of most baseline models has dropped significantly. 
For example, the NL2Code MRR of UniXcoder dropped sharply from 58.83\% to 15.88\%, likely due to unseen programming languages and the dataset scale.
By contrast, \Ours still maintains a high MRR of 57.76\%, outperforming the second-best model, TE3L, by 15.97\%.

\begin{table}[h!]
\centering
\caption{Performance in Cross-Language Scenarios on the Empirical Benchmark
}
\label{tab:retrieval_cross_ours}
\small
\vspace{-0.3cm}
\resizebox{\columnwidth}{!}{
\begin{tabular}{c rr rr rr rr}
\toprule
\multirow{2}{*}{\textbf{ Strategy }} & \multicolumn{2}{c}{\textbf{Java2Java}} & \multicolumn{2}{c}{\textbf{Python2Python}} & \multicolumn{2}{c}{\textbf{Java2Python}} & \multicolumn{2}{c}{\textbf{Python2Java}}\\
\cmidrule(lr){2-3} \cmidrule(lr){4-5} \cmidrule(lr){6-7} \cmidrule(lr){8-9}
& \textbf{MRR} & \textbf{MAP} & \textbf{MRR} & \textbf{MAP} & \textbf{MRR} & \textbf{MAP} & \textbf{MRR} & \textbf{MAP} \\
\midrule
Code2Code&83.53&66.04&92.12&82.60&85.26&77.82&91.92&78.16\\
Remix&90.59&71.88&93.97&84.46&91.17&83.31&92.99&79.16\\
Concat&87.04&69.33&92.88&83.51&89.81&81.62&92.27&78.34\\
Weight&89.00&69.81&93.29&82.73&91.36&82.77&92.49&78.41\\
\bottomrule
\end{tabular}
}
\vspace{-0.3cm}
\end{table}

Table \ref{tab:retrieval_cross_ours} illustrates the cross-language retrieval performance of \Ours on the empirical benchmarks. 
A comparison with Table \ref{tab:retrieval_cross} demonstrates UniCoR has strong cross-language retrieval capability, achieving average improvements of 39.09\% in MRR and 47.94\% in MAP. 
We further observe that when querying with Java using \Ours, both MAP and MRR decrease simultaneously when switching from the cross-language to the single-language scenario. 
This contrasts with the trend observed in Table \ref{tab:retrieval_cross}, where MAP decreases but MRR increases under the same condition. 
The alignment of MAP and MRR in our experiment demonstrates \Ours's effectiveness in minimizing the gap between them.

\begin{figure}[h]
    \centering
    \vspace{-0.3cm}
    \includegraphics[width=0.9\linewidth]{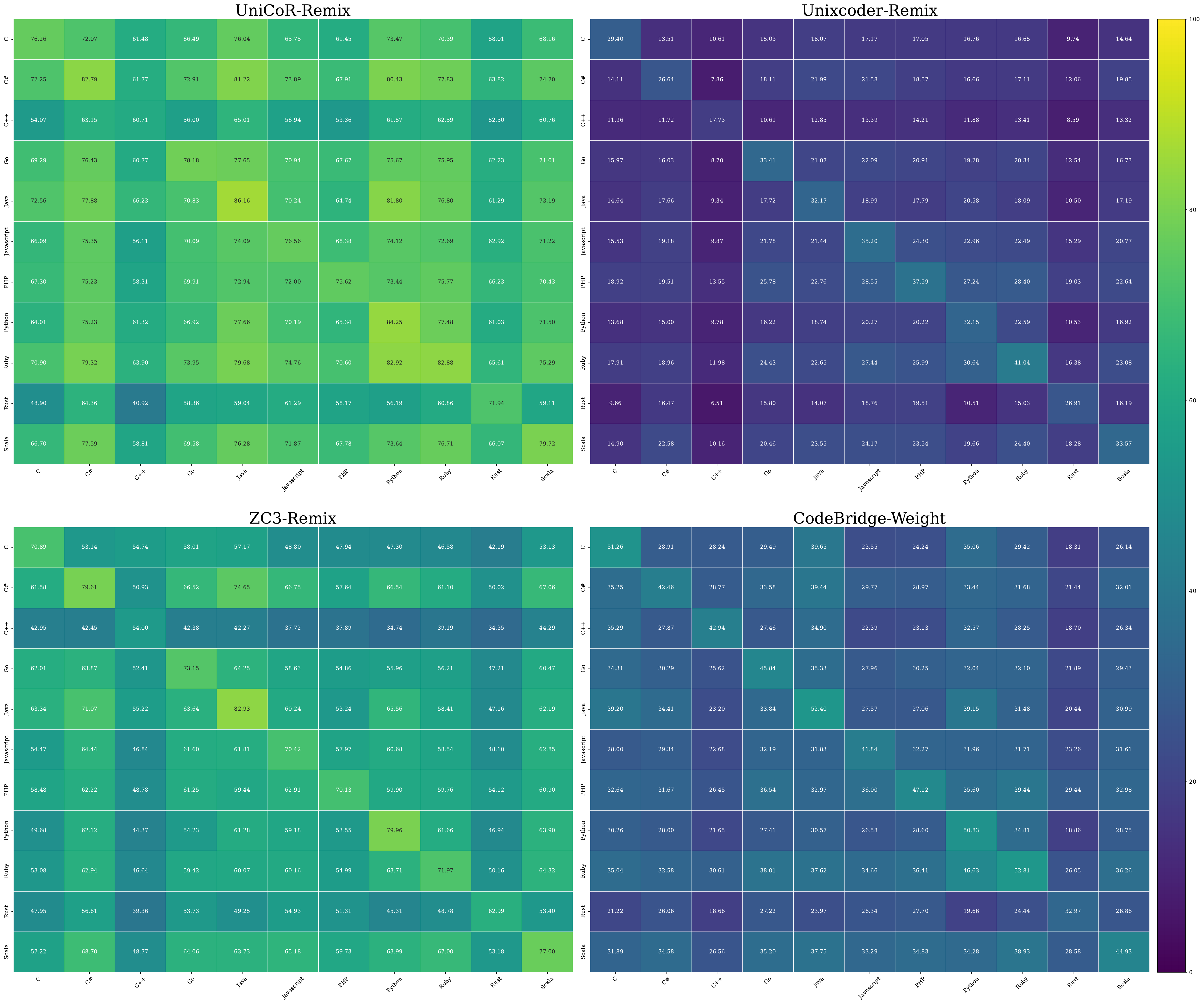}
    \vspace{-0.3cm}
    \caption{Model Comparison on Cross-Language Code Retrieval ( MRR ). The colors transition from dark to bright yellow, representing the MRR scores from low to high.}
    \vspace{-0.3cm}
    \label{fig:cross-language code retrieve visual}
\end{figure}

Figure \ref{fig:cross-language code retrieve visual} presents heatmaps that compare the MRR scores of several models for hybrid code retrieval across 11 programming languages on the XCodeEval benchmark. 
The rows and columns correspond to the source and target languages, respectively.
Further visual analysis confirms the cross-language advantage of \Ours.
Its heatmap shows large, uniform, and bright areas representing high MRR scores across all 11 languages, 
achieving an average improvement of 38.37\% for known languages and 32.92\% for unknown languages.
This demonstrates that \Ours learns a generalised, language-agnostic code representation. 
Meanwhile, the heatmaps for other models exhibit dark colors, reflecting their linguistic limitations and poor generalisation.

\begin{center}
    \fcolorbox{black}{green!5}{\parbox{0.98\linewidth}{
    \textbf{Answer to RQ3:} \Ours significantly outperforms all baseline methods across all retrieval strategies, achieving an average improvement of 
    8.64\% in MRR and 11.54\% in MAP. 
    Furthermore, it also exhibits excellent cross-language robustness, even on unseen programming languages during training.
    }}
\end{center}

\subsection{RQ4: How effective are the key designs of \Ours, and in what aspects are its strengths demonstrated?}
To answer RQ4, we investigate the internal mechanisms of \Ours from six perspectives to validate its design and effectiveness in order to address core retrieval challenges.

\begin{table}[h!]
\centering
\caption{The results of ablation studies( p-value < 0.05 )}
\label{tab:ablation}
\vspace{-0.3cm}
\resizebox{\columnwidth}{!}{
\begin{tabular}{l cc cc cc cc cc cc}
\toprule
\multirow{2}{*}{\textbf{Model}} & \multicolumn{2}{c}{\textbf{NL2Code}} & \multicolumn{2}{c}{\textbf{NL2NL}} & \multicolumn{2}{c}{\textbf{Code2Code}} & \multicolumn{2}{c}{\textbf{Remix}} & \multicolumn{2}{c}{\textbf{Concat}} & \multicolumn{2}{c}{\textbf{Weight}} \\
\cmidrule(lr){2-3} \cmidrule(lr){4-5} \cmidrule(lr){6-7} \cmidrule(lr){8-9} \cmidrule(lr){10-11} \cmidrule(lr){12-13}
& \textbf{MRR} & \textbf{MAP} & \textbf{MRR} & \textbf{MAP} & \textbf{MRR} & \textbf{MAP} & \textbf{MRR} & \textbf{MAP} & \textbf{MRR} & \textbf{MAP} & \textbf{MRR} & \textbf{MAP} \\
\midrule
UniXcoder&58.83&36.98&84.74&64.20&68.93&46.70&70.88&47.72&74.37&50.29&75.38&50.59\\
\cmidrule(r){2-13}
\quad w/o Data&65.47&46.96&78.18&58.18&77.3&58.4&78.08&58.21&78.09&58.78&79.71&60.46 \\
\quad  w/o MPCL&79.58&66.14&87.9&75.71&87.88&76.13&90.98&78.73&89.13&76.97&90.56&77.96 \\
\quad  w/o RDCL&80.46&67.85&89.01&77.75&88.35&77.81&91.53&80.12&90.18&78.98&91.44&79.9 \\

\quad +\Ours&81.81&70.24&89.01&78.88&89.12&78.75&92.26&81.55&90.81&80.20&92.09&80.84\\

\midrule
CodeBERT&7.80&3.64&53.56&24.86&24.88&10.41&20.33&8.52&15.55&6.84&24.97&10.44
\\
\quad +\Ours&60.33&47.42&75.74&60.43&76.31&62.54&75.14&60.07&72.68&58.54&78.01&63.77
\\
\midrule
GraphCodeBERT&28.84&14.74&74.91&49.06&50.73&29.19&51.50&29.09&50.98&28.89&53.57&30.58\\
\quad +\Ours&79.79&67.83&88.53&78.12&88.31&77.83&91.45&80.68&89.25&78.64&90.59&79.5
\\
\bottomrule
\end{tabular}
}
\vspace{-0.1cm}
\end{table}

(1) \textbf{Ablation Studies}: 
We quantify the contribution of each key component in \Ours through ablation studies. 
As shown in Table \ref{tab:ablation}, the full version of \Ours (using UniXcoder as the backbone) achieves the best performance. 
\textbf{Removing any single component leads to a decline in performance, which highlights the necessity of each component and the integrity of the overall framework.}
Furthermore, quantitative analyses confirm that each component contributes statistically significant and consistent gains (p-value < 0.05).
\textbf{w/o Data}: Removing the functionally equivalent data augmentation strategy leads to the most significant performance degradation. 
This demonstrates that high-quality, diverse positive samples are essential for learning deep semantic equivalence. 
Interestingly, this ablated configuration is equivalent to fine-tuning the original model with the original training data while retaining the Representation Distribution Consistency Learning (RDCL), and we observe that this still effectively improves the Code2Code strategy.
\textbf{w/o MPCL}: The elimination of the intra-modal contrastive objectives (NL-NL and Code-Code) from the Multi-Perspective Contrastive Learning (MPCL) framework also results in a substantial decline in performance. 
This finding suggests that ensuring semantic consistency within each modality is paramount for establishing a robust cross-modal alignment.

(2) \textbf{Generalizability Studies}: 
The generalizability of \Ours is evaluated through its application to two alternative backbone models: CodeBERT and GraphCodeBERT.
As demonstrated in Table \ref{tab:retrieval_final} (in the CodeBERT row), the baseline CodeBERT model attains a mere 7.80\% in MRR on the NL2Code task. 
The application of \Ours during training has been shown to significantly enhance performance, resulting in an absolute improvement of over 52.5\%, with an MRR score of 60.33\%.
This demonstrates that \textbf{\Ours is a highly general and model-agnostic training framework that can significantly enhance the representation capabilities of pre-trained code models}.

\begin{figure}[htbp]
    \centering
    \begin{subfigure}{0.49\columnwidth}
        \centering
        \includegraphics[width=\linewidth]{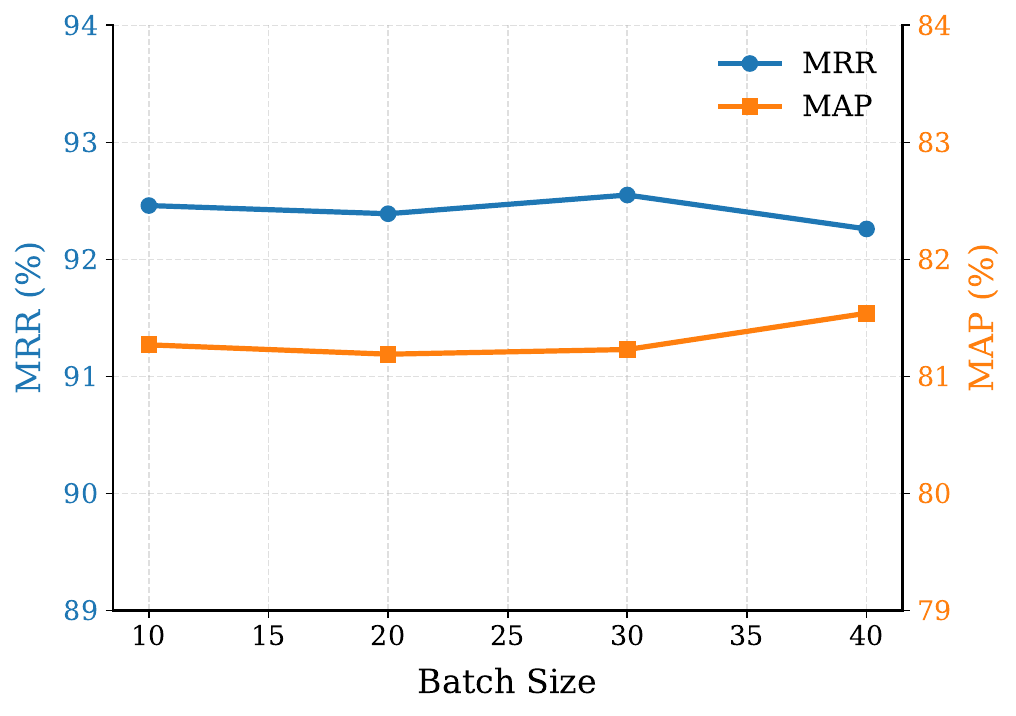}
        \caption{Batch Size}
        \label{fig:sub-a}
    \end{subfigure}
    \hfill 
    \begin{subfigure}{0.49\columnwidth}
        \centering
        \includegraphics[width=\linewidth]{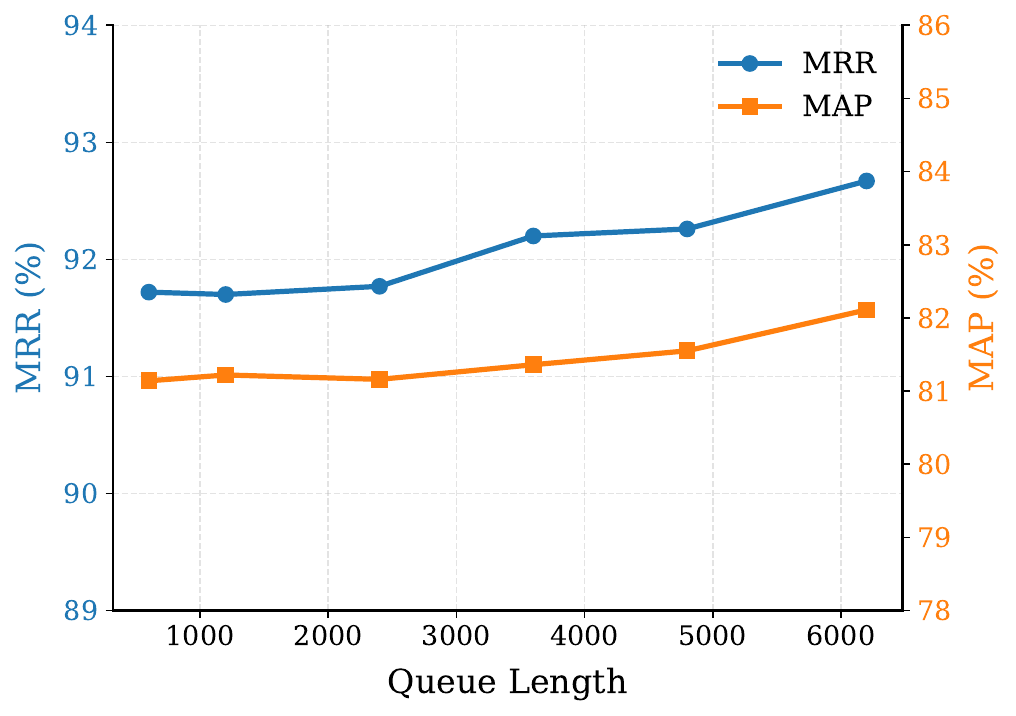}
        \caption{Queue Size}
        \label{fig:sub-b}
    \end{subfigure}
    \vspace{-0.1cm}
    \caption{The impact of different hyperparameters on Remix-based hybrid code retrieval}
    \label{fig:side-by-side}
    \vspace{-0.3cm}
\end{figure}

(3) \textbf{Hyperparameter Sensitivity Analysis}:
To explore the sensitivity of model performance to key hyperparameters, we conduct an analysis of batch size and queue size.
As demonstrated in Figure \ref{fig:sub-a}, the performance remains consistent when the batch size increases from 10 to 40, with fluctuations not exceeding 0.4\%.
As demonstrated in Figure \ref{fig:sub-b}, as the queue size increases from 600 to 6200, both MRR and MAP exhibit only a marginal upward trend (0.82\% and 0.92\%, respectively). 
The findings indicate that \textbf{\Ours exhibits minimal sensitivity to the key hyperparameters}.

\begin{table}[h]
\centering
\small
\caption{The weight report of Weighted Fusion-based hybrid code retrieval}
\label{tab:retrieval_hybrid_weight}
\vspace{-0.3cm}
\resizebox{0.85\columnwidth}{!}{
\begin{tabular}{cccccc}
\toprule
\textbf{Model} & \textbf{Weight-NL2Code} & \textbf{Weight-Code2Code}\\
\midrule
CodeBERT & 7.0($\pm$ 1.2) & 93.0($\pm$ 1.2)\\
GraphCodeBERT & 47.0($\pm$ 9.7) & 53.0($\pm$ 9.7)\\
UniXcoder & 47.0($\pm$ 6.1) & 53.0($\pm$ 6.1)\\
BGE& 36.0($\pm$ 4.9) & 64.0($\pm$ 4.9)\\
ZC3& 33.0($\pm$ 0.8) & 67.0($\pm$ 0.8)\\
CoCoSoDa& 48.0($\pm$ 3.8) & 52.0($\pm$ 3.8)\\
CodeBridge& 19.0($\pm$ 0.9) & 29.0($\pm$ 9.9)\\
\midrule
\Ours& 48.0($\pm$ 4.4) & 52.0($\pm$ 4.4)\\
\bottomrule
\end{tabular}
}
\vspace{-0.1cm}
\end{table}

(4) \textbf{Modality Co-Analysis:} In order to verify whether \Ours mitigates the "modality imbalance" problem identified in earlier empirical studies, we analyzed the optimal weights in the Weighted strategy (See Table \ref{tab:retrieval_hybrid_weight}).
The optimal weights for the NL-Code and Code-Code perspectives in \Ours are 48\% and 52\%, respectively, and their contributions are relatively balanced. 
In comparison with the backbone UniXcoder, \Ours attains a more balanced utilisation of modalities while concomitantly enhancing stability, thereby reducing the standard deviation of the optimal weight across datasets from 6.07 to 4.45.
This strongly indicates that \textbf{\Ours successfully learns equally strong and reliable representations across modalities, achieving robust modality collaboration.}

\begin{table}[h!]
\centering
\small
\caption{Inference Time Comparison on the CodeSearchNet Dataset}
\vspace{-0.3cm}
\resizebox{0.9\columnwidth}{!}{
\begin{tabular}{ccc}
\toprule
\textbf{Model} & \textbf{Total Inference Time (s)} & \textbf{Embedding Time (s)}\\
\midrule
Roberta       & 0.010 & 0.009 \\ 
CodeBERT      & 0.010 & 0.009 \\ 
GraphCodeBERT & 0.010 & 0.009 \\ 
UniXcoder     & 0.010 & 0.009 \\ 
ZC3           & 0.010 & 0.009 \\
\midrule
BGE       & 0.027 & 0.026 \\
BM25      & 0.298 & -     \\
TE3L       & 0.327 &0.325 \\
SGPT      & 0.152 & 0.151 \\
DS2Vec    & 0.159 & 0.158 \\
Qwen2Vec  & 0.151 & 0.150 \\
\midrule
UniCoR        & 0.010 & 0.009\\
\bottomrule
\end{tabular}
\label{tab:time_performance}
}
\vspace{-0.3cm}
\end{table}

(5) \textbf{Inference Time:} To evaluate the practical efficiency of \Ours, we conduct an average inference-time analysis on the CodeSearchNet dataset using 3,100 queries.
As shown in Table \ref{tab:time_performance}, \Ours processes each query in 0.010 s (0.009s for embedding generation) which is comparable with the fastest model. 
These results demonstrate that \Ours maintains comparable computational efficiency to existing transformer-based encoders while delivering enhanced retrieval performance, confirming its suitability for large-scale, latency-sensitive code retrieval applications.

\begin{figure}
    \centering
    \includegraphics[width=0.89\linewidth]{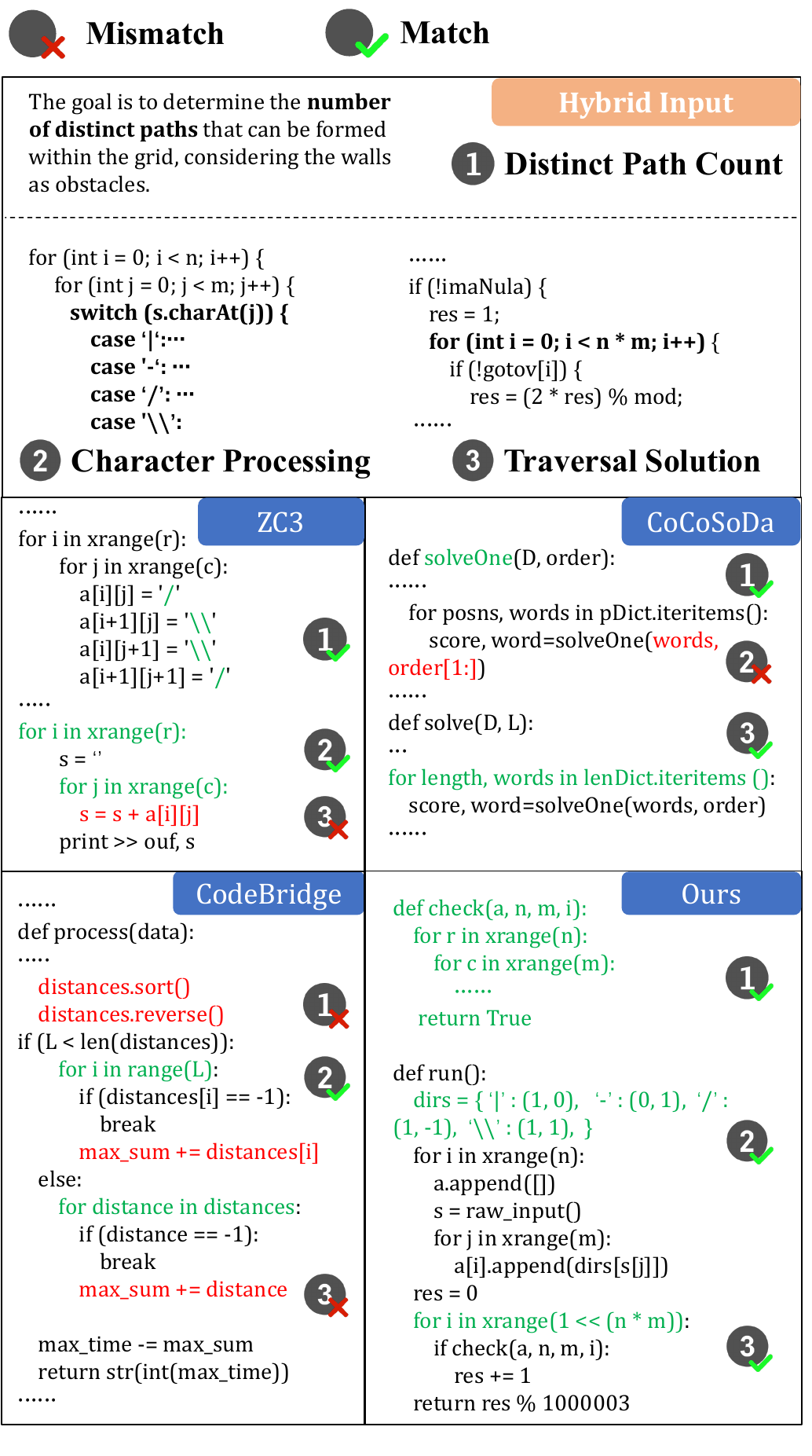}
    \vspace{-0.3cm}
    \caption{Case Study on Problem Retrieval via Hybrid Code Retrieval}
    \label{fig:case}
    \vspace{-0.3cm}
\end{figure}

(6) \textbf{Case Study:} 
A case study further illustrates these capabilities (See Fig.\ref{fig:case}). 
This case involves a hybrid query that contains a natural language description and Java implementation for "counting the number of unique paths in a character grid".
The objective is to identify and retrieve equivalent Python implementations.
Success requires understanding three core functional requirements: 1) processing character grid data; 2) using a traversal algorithm; 3) ensuring path uniqueness. 
Baseline models such as UniXcoder and ZC3 can match surface-level features, including the "character grid" and "traversal" features. However, the retrieved functions were found to be functionally inconsistent with the objective of the query (e.g., modifying the grid rather than counting paths).
We find that other models, including CodeBridge and CoCoSoDa, identify similar patterns.
However, these models fail to satisfy the requirement of processing a character grid.
In contrast, \Ours successfully retrieved a functionally equivalent Python implementation.
This clearly shows that \textbf{\Ours has the capacity for seamless integration of high-level functional intent derived from natural language with the underlying algorithmic logic of code, thereby facilitating accurate hybrid retrieval.}

\begin{center}
    \vspace{-0.2cm}
    \fcolorbox{black}{green!5}{\parbox{0.98\linewidth}{
    \textbf{Answer to RQ4:} 
    Experiments confirm that our success stems from our robust design.
    Ablation studies verify the necessity of each component, particularly the data augmentation strategy.
    \Ours effectively mitigates modality imbalance, achieving an almost equal contribution of 48\%/52\% from the NL and Code perspectives.
    Finally, \Ours is robust, demonstrating less than 1\% fluctuation in performance across a wide range of key hyperparameters.
    }}
\end{center}

%% file: RelationWork.tex
\section{Related Work}
\label{sec:relatedwork}

Code retrieval aims to find relevant code snippets from a codebase given a query. 
Early code retrieval techniques, such as Sourcerer \cite{Sourcerer} and Codehow \cite{Codehow}, relied on information retrieval methods including lexical matching and structural feature analysis. 
However, these methods were susceptible to tokens of codes and queries and could not understand the deep semantics. 
The advent of deep learning has led a paradigm shift.
Foundational research, such as DeepCS \cite{Deepcodesearch} and Code2Vec \cite{PACMPL2019Code2vec}, represented code snippets as fixed-length, dense semantic vectors, and laid the groundwork for subsequent research.
Following research further enriched these representations by leveraging graph structures to capture more structural information \cite{GNNcoder1, GNNcoder2, Gnncoder3} or by mining primitive program logic from intermediate representations like LLVM \cite{ICML2021OSCAR}.

Motivated by the success of pre-trained language models in natural language processing, methods based on pre-trained language models such as CodeBERT \cite{feng2020codebert}, GraphCodeBERT \cite{guo2020graphcodebert}, and UniXcoder \cite{guo2022unixcoder} have become widely adopted approaches for code representation. 
To further enhance the code representation capability, contrastive learning has been widely used for code retrieval. 
Various strategies were proposed to enhance alignment between query and code, which included: 
1) Constructing Various Positive Samples. 
This process involved the generation of multiple views of the same code, with the objective of enhancing representation robustness. 
For instance, RFMC-CS \cite{JASE2025RFMCCS} paired source code with its graph structures to form positive examples, while CoCoSoDa \cite{Shi2023CoCoSoDA} used soft data augmentation to generate diverse positive samples. 
2) Hard Negative Mining. This approach focused on constructing challenging negative samples that compelled the model to learn more fine-grained semantic differences \cite{TOSEM2025CoCoHaNeRe, li2025scodesearcher, ICSE2025Hedgecode}. 

Beyond general code retrieval, another research has focused on Code2Code and cross-domain retrieval settings.
In the Code2Code setting, prior studies have generally followed two main branches.
The first branch converts code into graphs \cite{TSE2023GOC, ALIP2023CCCS} or assembly language \cite{ICSE2024Prism} to mitigate syntactic differences.
The second branch focuses on aligning representations, evolving from dependent parallel corpus-based supervised learning \cite{ICPC2022C4} to unsupervised methods employing adversarial training and recurrent consistency \cite{AAAI2024AdaCCD, li2023zc3}.
In the context of cross-domain code retrieval, research has explored the potential of meta-learning \cite{ICSE2022CDCS} to enhance model transferability. 
These studies have explored the use of generative models to synthesise queries for fine-tuning \cite{TOSEM2023RAPID}, and the integration of LLMs to generate multi-perspective representations and facilitate zero-shot fusion \cite{liang2025codebridge}.
\textbf{The models chosen in our experiments are state-of-the-art in their respective domains and are intended to represent the latest advancements in the field.}

The proposed framework, \Ours, differs from existing code retrieval approaches in both its design and methodology.  First, from a design perspective, \Ours can easily integrate existing code LLMs and improve their performance on code retrieval. Second, from a methodology perspective, we design a multi-perspective supervised contrastive learning module to learn unified semantics and achieve modality interaction. Lastly, to enhance cross-language generalization, we incorporate a representation distribution consistency learning module to capture the language-agnostic logical essence of code. 

%% file: Discussion.tex
\section{Conclusion and Future Work}
\label{sec:conclusion}
In this paper, we explore the challenging and underexplored problem of hybrid and cross-language code retrieval. 
We first conduct a comprehensive empirical study and identify three key challenges that the pretrained code models face in this field.
To overcome these problems, we introduce \Ours. 
Extensive experiments on two benchmarks show that \Ours achieves state-of-the-art performance in hybrid code retrieval, achieving an average improvement of 8.64\% in MRR and 11.54\% in MAP, while also substantially improving zero-shot cross-language generalization.
In the future, we plan to integrate our work into LLM-based development tools, such as a function in MCP, so that more developers can utilize our research.

\vspace{-0.3cm}
\section*{Acknowledgments}
The authors would like to thank the anonymous reviewers for their insightful comments.
This work has been supported by the National Natural Science Foundation of China under Grant No. 62472447, the Hunan Provincial Natural Science Foundation of China under Grant No. 2024JK2006, the Science and Technology Innovation Program of Hunan Province under Grant No. 2023RC1023, and the Hunan Provincial Graduate Research and Innovation Project under Grant No. CX20250244. This work was carried out in part using computing resources at the High Performance Computing Center of Central South University.

\vspace{-0.3cm}
\section*{Data Availability}

The data and the source code are available in the repository \url{https://github.com/css518/UniCoR}.